%% file: kddads2019_main.tex
\documentclass[sigconf]{acmart}

\def\BibTeX{{\rm B\kern-.05em{\sc i\kern-.025em b}\kern-.08emT\kern-.1667em\lower.7ex\hbox{E}\kern-.125emX}}
    
\acmPrice{15.00}
\acmDOI{10.475/123_4}

\acmISBN{123-4567-24-567/08/06}

\acmConference[KDD'19]{KDD Workshop on Applied Data Science for Healthcare}{August 2019}{Anchorage, Alaska USA}
\acmYear{2019}
\copyrightyear{2019}

\acmArticle{4}
\acmPrice{15.00}

\setcopyright{acmlicensed}
\usepackage{graphicx}
\usepackage{amsmath}
\usepackage{amssymb}
\usepackage[compact]{titlesec}
\titlespacing{\section}{0pt}{2ex}{1ex}
\titlespacing{\subsection}{0pt}{1ex}{0ex}
\titlespacing{\subsubsection}{0pt}{0.5ex}{1ex}

\usepackage{xcolor}
\usepackage{mathtools}
\usepackage{caption}
\usepackage{marginnote}
\usepackage{hhline}% http://ctan.org/pkg/hhline
\usepackage{xspace}
\usepackage{enumitem}

\usepackage[colorinlistoftodos, obeyDraft]{todonotes}
\usepackage{tikz}
\usetikzlibrary{shapes.geometric}
\usetikzlibrary{arrows.meta,arrows}

\usepackage{caption}
\usepackage[colorinlistoftodos]{todonotes}

\usepackage{eso-pic, rotating, graphicx}
\begin{document}

\title{SLAM Endoscopy enhanced by adversarial depth prediction}

\author{Richard J. Chen}
\authornote{Contributed equally.}
\affiliation{%
  \institution{Johns Hopkins University}
}
\email{rchen40@jhu.edu}

\author{Taylor L. Bobrow}
\authornotemark[1]
\affiliation{%
  \institution{Johns Hopkins University}
}
\email{tbobrow1@jhu.edu}

\author{Thomas Athey}
\affiliation{%
  \institution{Johns Hopkins University}
}
\email{tathey1@jhmi.edu}

\author{Faisal Mahmood}
\affiliation{%
  \institution{Johns Hopkins University}
}
\email{faisalm@jhu.edu}

\author{Nicholas J. Durr}
\affiliation{%
  \institution{Johns Hopkins University}
}
\email{ndurr@jhu.edu}

\begin{abstract}
Medical endoscopy remains a challenging application for simultaneous localization and mapping (SLAM) due to the sparsity of image features and size constraints that prevent direct depth-sensing. We present a SLAM approach that incorporates depth predictions made by an adversarially-trained convolutional neural network (CNN) applied to monocular endoscopy images. The depth network is trained with synthetic images of a simple colon model, and then fine-tuned with domain-randomized, photorealistic images rendered from computed tomography measurements of human colons. Each image is paired with an error-free depth map for supervised adversarial learning. Monocular RGB images are then fused with corresponding depth predictions, enabling dense reconstruction and mosaicing as an endoscope is advanced through the gastrointestinal tract.  Our preliminary results demonstrate that incorporating monocular depth estimation into a SLAM architecture can enable dense reconstruction of endoscopic scenes. %We experimentally validate our approach using endoscopy videos of phantom and porcine colon tissue with ground truth from a commercial 3D scanner.
\end{abstract}

%
% The code below should be generated by the tool at
% http://dl.acm.org/ccs.cfm
% Please copy and paste the code instead of the example below.
%



\maketitle
\renewcommand{\shortauthors}{RJ. Chen et al.}

\input{kddads2019_body}

\bibliographystyle{ACM-Reference-Format}

\clearpage

% please don't  input invalid LaTeX
%\input{kdd_optional_reproducibility.tex}

\end{document}

%% file: kddads2019_body.tex
\section{Introduction}
Colonoscopy screening is the standard of care for detecting and diagnosing gastrointestinal conditions, with 15 million colonoscopies performed annually in the United States [1]. Still, 50,000 deaths are expected to occur in the United States in 2019 from Colorectal Cancer (CRC), making it the second deadliest cancer variety in the United States [2]. The disparity in caregivers' ability to detect colorectal lesions is a cause for concern, as an estimated 20\% of lesions go undetected during routine screenings [3], with a strong correlation existing between the examining physician's adenoma detection rate (ADR) and the patient's likelihood of experiencing advanced stage or fatal interval cancer - the diagnosis of cancer between routine screenings [4]. As a minimally invasive surgical (MIS) procedure, scene understanding in colonoscopy is inherently challenging, as the examining physician is tasked with evaluating a patient's gastrointestinal system using only localized, monocular images with little ability to aggregate features as the procedure progresses. For example, Cecal Intubation Rate (CIR), or percent of procedures with confirmed imaging from rectum to cecum, is considered a key performance indicator for a successful exam, yet positively identifying the cecum is often difficult with only monocular images to rely on [5]. Routine colonoscopy may benefit from the aggregation, or mosaicing, of image information to form a more complete picture when making colonoscopic assessments.

\begin{figure*}
\centering
\includegraphics[width=\textwidth]{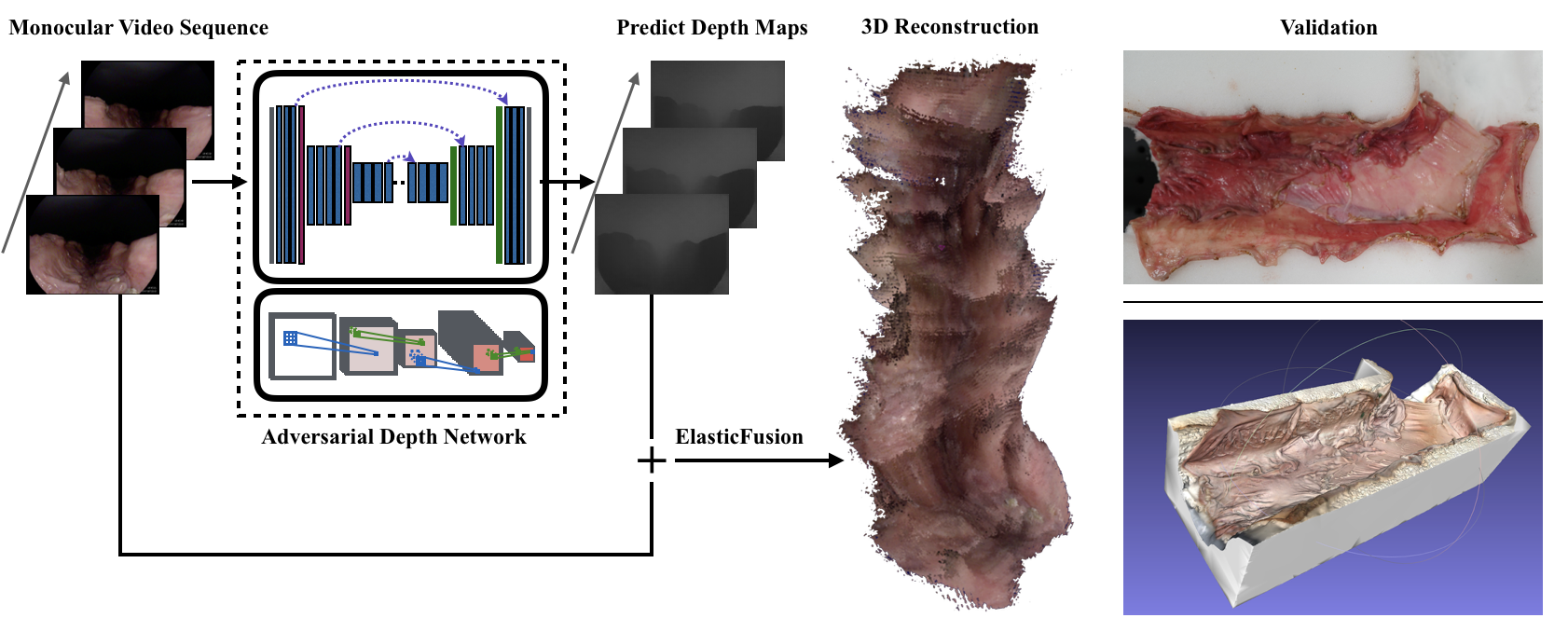}
\caption{\textmd{Our framework for SLAM-endoscopy. We first use Siemens VRT technology to create photorealistic training data (cinematic renderings) for monocular depth estimation. We then use adversarial training to incorporate context-aware information in our network to accurately predict depth, which we finally fuse with RGB into ElasticFusion to create a dense surfel point cloud of the GI.}}
\end{figure*}

Despite the advancements made in computer vision and visual scene understanding, optical colonoscopy remains a challenging environment with conventional endoscopes bolstering wide field of view (FOV) monocular detectors and a wide range of working distances. With the emergence of simultaneous localization and mapping (SLAM) [6] for scene reconstruction came its application to the endoscopic setting, with the first work by Montiel et. al. who developed a monocular SLAM approach to sparsely recover the 3D geometry of the abdominal cavity [7]. More recently, the landscape of SLAM endoscopy has been focused around extending feature-based SLAM systems such as ORB-SLAM with depth estimated from stereo cameras to track and map more features [8,9,10,11]. These feature-based SLAM systems tend to work well in rigid, well lit scenes with large working distances, but can fail to track or generate sufficient features for dense reconstructions in settings such as colonoscopy. The paucity of distinguishing features, tissue homogeneity, deforming surface, and highly variable specular appearance of the lumen can cause inconsistencies in estimating camera pose for systems such as ORB-SLAM, as not enough ORB features can be reliably tracked. Moreover, many of the strategies proposed above do not satisfy the hardware limitations of endoscopes maintaining a monocular camera source and wide field-of-view. Because of these issues, monocular depth estimation is a challenging, yet critical problem that must be solved to enable RGB-D SLAM systems to reconstruct scenes in the GI. %with many current approaches limited by diversity of annotated data.

In this work, we present a strategy for monocular depth estimation in endoscopy that is robust to specular reflection, and a framework for performing SLAM-endoscopy by fusing monocular RGB images with corresponding depth predictions (Figure 1). We offer one of the first dense reconstruction of the gastrointestinal tract using only monocular images, with phantom and ex-vivo tissue models paired with ground truth for qualitative assessment, with quantitative assessment pending in work-in-progress.

\section{Method}

\subsection{Monocular Depth Estimation using Conditional GANs}
Monocular depth estimation is a challenging problem in endoscopy, with current approaches suffering from poor generalization performance due to lack of diverse training data, overfitting to patient-specific textures and colors, and fail to incorporate non-local information for learning depth cues. To overcome these issues, we develop an adversarial approach for context-aware monocular depth estimation that is able to generalize to unknown modes of patient data, shown in Figure 2 [12]. We denote $A$ and $B$ as the RGB and depth image domains respectively, $G$ as a mapping function $G: A\rightarrow B$ that maps RGB to depth, and $D$ as the discriminator network for $G$.

\subsubsection{Conditional GAN Objective:}
The conditional Generative Adversarial Network (GAN) framework consists of two networks that compete against each other in a \textit{min-max} game to respectively minimize and maximize the objective, $\text{min}_G \text{max}_{D} \mathcal{L}(G, D)$. The generator $G$ is our depth estimation network that learns a mapping from $A$ to $B$, and the discriminator $D$ distinguishes between real and synthesized pairs of depth and RGB. We can use this framework for pixel-wise depth estimation by mixing the adversarial loss term $\mathcal{L}_{\text{GAN}}$ with a per-pixel loss term  $\mathcal{L}_\text{L1}$ to penalize both the joint configuration of pixels and accuracy of the estimated depth maps.  In the mapping $G: A \rightarrow B$, the $L_1$ loss term is used to score the accuracy of the depth estimation by $G$, with its strength controlled by $\lambda$. We can express the adversarial objective as the binary cross entropy loss of $D$ in classifying real/synthesized pairs. The motivation for using an adversarial loss term is to incorporate non-local information for important depth cues, such as the the inverse square fall-off in light intensity with propagation distance from the light source. This non-local loss information is calculated by the discriminator, which classifies overlapping pairs of image and depth patches as being real or synthetic as the adversarial loss. By controlling the size of the patch, we can control how much global / non-local information to include in the image. In total, we can express these loss terms as:

\begin{equation}\label{eq:Tk}
  \begin{aligned}
 \mathcal{L}_{\text{GAN}}(G, D) = \mathbb{E}[\text{log}D(A,B)] +\\ \mathbb{E}[\text{log}(1-D(B,G(A))]
  \end{aligned}
\end{equation}

\begin{equation}\label{eq:Tk}
  \begin{aligned}
 \mathcal{L}_{\text{L1}}(G) = \mathbb{E}[||B-G(A)||_1] 
  \end{aligned}
\end{equation}

\begin{equation}\label{eq:Tk}
  \begin{aligned}
    \text{min}_G \text{max}_{D} \mathcal{L}_{\text{GAN}}(G,D) + \lambda\mathcal{L}_{\text{L1}}(G)
  \end{aligned}
\end{equation}

\begin{figure}
\centering
\includegraphics[width=\columnwidth]{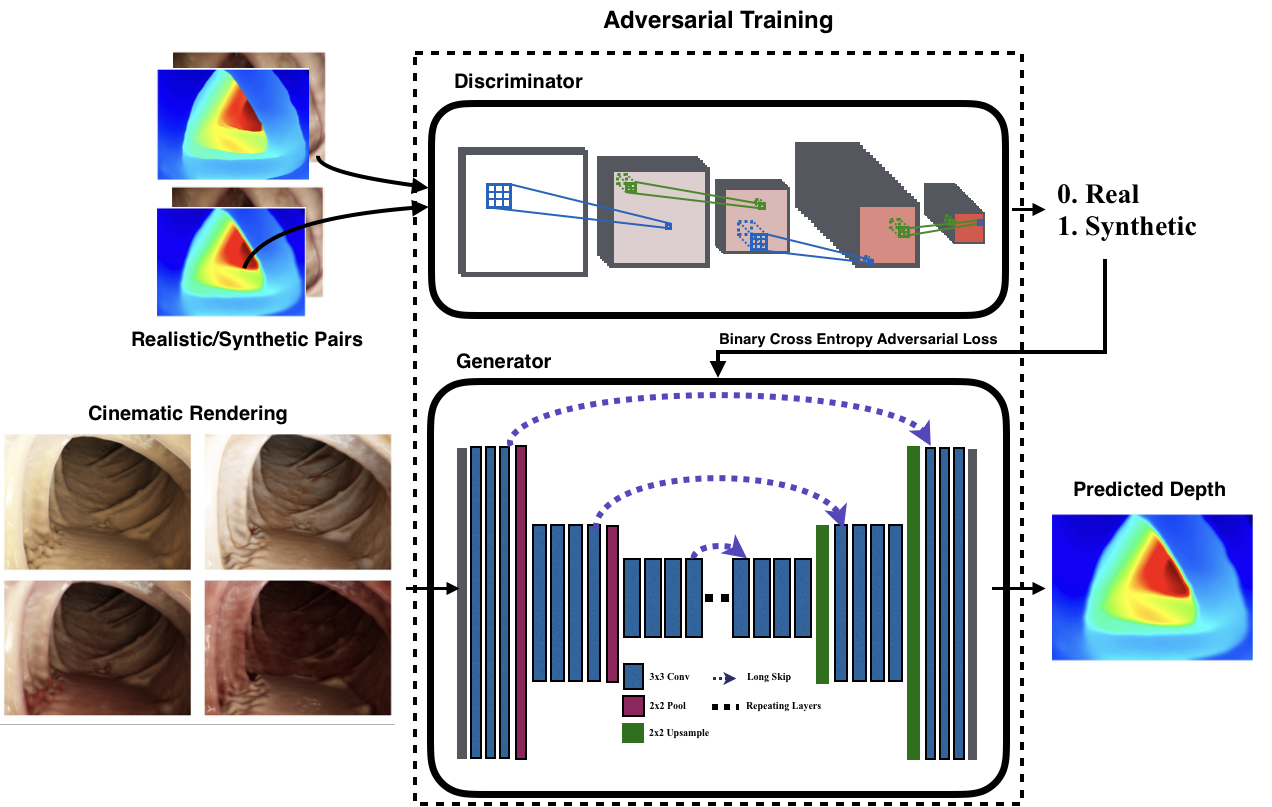}
\caption{\textmd{Adversarial framework for monocular depth estimation.}}
\end{figure}

\subsubsection{Model \& Training Details:}
Encoder-decoder networks, such as the U-Net, are commonly used in many deep network approaches for monocular depth estimation. The U-Net architecture draws skip connections between convolution layers on the encoder path and up-sampling layers on the decoder path that have the same spatial size. To stabilize the GAN training procedure of the U-Net as the generator, we applied spectral normalization to convolution layers, which controls the spectral norm $\sigma$ of the convolution weights $W$ in the network would be bounded by the Lipschitz constraint $\sigma(W) = 1$. We also further stabilize the discriminator by using a buffered data input from the generator, which consists of previously generated and classified pairs and ground truth data. In our observations, we found that these techniques reduced visual artifacts made by the generator, which resulted in more smoothly-varying depth estimates.

\subsubsection{Training Data:} The Cinematic VRT technology developed by Siemens Healthcare uses a novel technique that can simulate light scattering and extinction through turbid media, creating natural and photorealistic 3D representation for medical scans that mimic the physical lighting experienced in real tissue [13]. We used this software to generate a diverse set of four renderings for 1190 endoscopic scenes, with 3570 scenes with annotated depth rendered in total. The CT colonoscopy data used was acquired from 13 patients from the NIH Cancer Imaging Archive (TCIA). By including renderings of the same colon image with different colors and textures in the training set, the network can learn more domain-invariant features, which would allow it to generalize well to other tissue models. Visualization and performance metrics of predicted depths using conditional GANS for endoscopic scenes are shown in Figure 3 and Table 1. A held-out set was created by holding out data from two patients from the training set, which approximately created a 80-20 split in training and validation samples.

\subsection{Dense Surface Reconstruction}
We fuse endoscopic monocular RGB video frames $(M)$ with estimated depth frames $D$ as input to ElasticFusion for surface reconstruction [14]. At a high level, ElasticFusion seeks to: i.) utilize pairs of RGB-D frames to compute a surface element (surfel) model of the imaged scene, ii.) test a registration between the current observable scene and the neighboring unobserved scene, and iii.) compare each acquired frame against a dictionary of predicted model views. If a match is found, the algorithm computes an alignment between the current local model and the entire global model. Specular reflection is detected through a light source pose estimation process in ElasticFusion, in which $M$ and $D$ is used to estimate a set of discrete light sources, which is then used to add rendered lighting in a scene.

\begin{figure}[!h]
\centering
\includegraphics[width=\columnwidth]{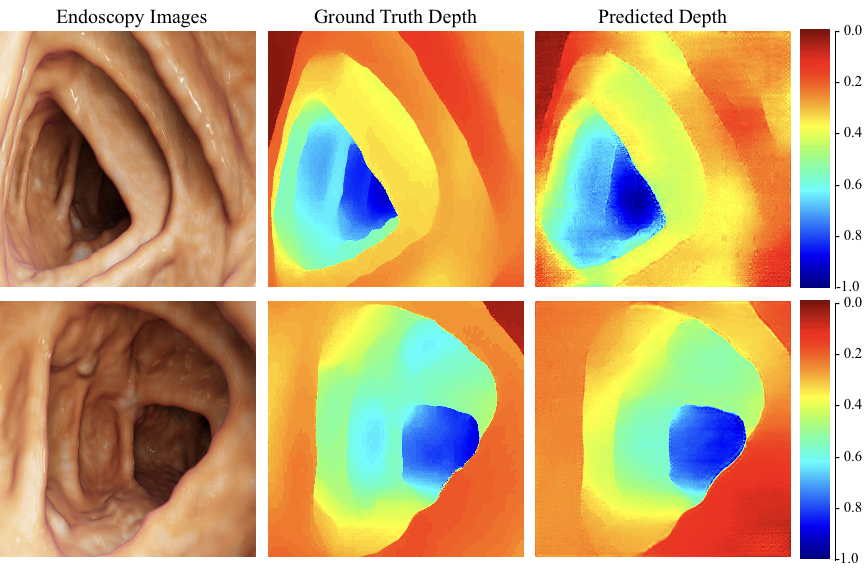}
\caption{\textmd{Cinematic rendering endoscopy images from a held-out set with corresponding ground truth depth and estimated depths. For visualization, distance is normalized between values 0 and 1, which correspond to closer and farther distances.}}
\end{figure}

\begin{table}[!h]
\begin{tabular*}{\columnwidth}{@{\extracolsep{\fill}} c|c c c}
\hline
Method & rel $\downarrow$& $log_{10} $ $\downarrow$& rms $\downarrow$ \\
\hline
U-Net-Adv. 			 & 0.312	& 0.012 & 0.054\\
\hline
\end{tabular*}
\caption{\textmd{Performance evaluation for cinematically rendered endoscopy images on the held-out set.}}
\end{table}

\iffalse
\begin{table}[!h]
%\setlength\tabcolsep{0pt}
\begin{tabular*}{\columnwidth}{@{\extracolsep{\fill}} c|c c}
\hline
Method & rms $\downarrow$ & Number of Points \\
\hline
Orb-SLAM      					 	 & - & 6495 \\
\textbf{Our Method} 			 & \textbf{0.4871} & \\
\hline
\end{tabular*}
\caption{\textmd{Quantitative Assessment on ex-vivo porcine colon tissue.}}
\end{table}
\fi

\subsection{Experimental Setup}

\begin{figure*}
\centering
\includegraphics[width=\textwidth]{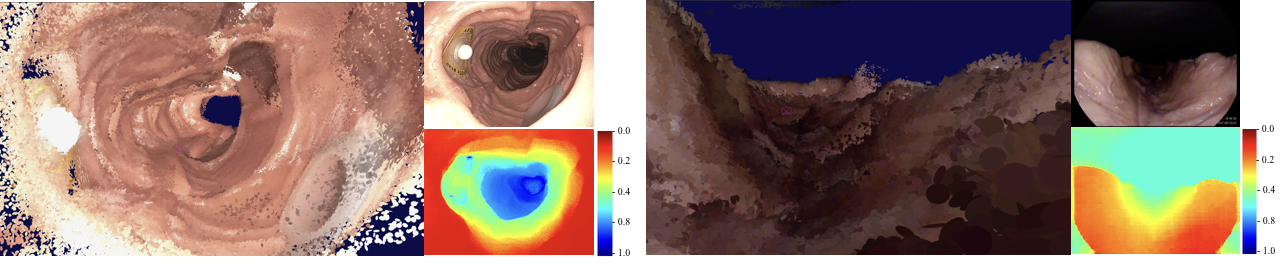}
\caption{\textmd{Qualitative Assessment on phantom tissue (left) and ex-vivo porcine colon tissue (right), with corresponding RGB frame and predicted depth measurement. 3D reconstruction made available here:  \href{https://youtu.be/7I-d5LwIAQI}{\color{blue} https://youtu.be/7I-d5LwIAQI}}}
\end{figure*}

To validate the application of ElasticFusion to colonoscopy, we recorded phantom and tissue models featuring realistic lumen topography for qualitative evaluation. A silicone circular colon phantom model (Colonoscopy Trainer 2003, the Chamberlain Group) was used, with a conventional Olympus colonoscope (Olympus Medical, CFHQ-190L) used to navigate. For demonstration with real gastrointestinal tissue, we utilized freshly harvested porcine colon for a realistic representation of human tissue. A 3 inch diameter half pipe was cut through the surface of a dense foam block to serve as a scaffold for the colon tissue. Ridges were cut into the walls of the foam in order to mimic the semilunar folds of the large intestine. The tissue was blanketed over the foam scaffold and pinned in to place. A conventional Pentax colonoscope (Pentax Medical, EC34-i10L) was mounted to a linear stage and passed through the tissue scaffold, capturing a video sequence similar to that which is captured in colonoscopy. %In this work-in-progress, To mitigate any tracking errors during 3D scanning of the ex-vivo model, the tissue was lightly sprayed with a Gold Bond spray. Fiduciary markers were pinned to the model to aide with manual registration, which was followed by iterative closest point (ICP) to closely align the model for evaluation. 

% A comparison of our Endoscopic SLAM framework with feature-based SLAM systems as used by relevant literature (ORB-SLAM) was done on qualitatively on both tissue models. A quantitative comparison was not feasible,   To the best of our ability, we used the ORB-SLAM system that has been adopted by many other endoscopy literature to obtain another 3D reconstruction for comparison. We use the root mean squared error (RMSE) of the reconstructions registered with the ground truth 3D scan, and the number of points tracked, as two metrics for evaluating accuracy of dense reconstruction.

\section{Results \& Discussion}

In this work, we present the first dense reconstruction of large phantom and real porcine colon tissue models of the GI tract, shown in Figure 4. In the phantom model, we were able to accurately reconstruct many of the shape and details of the circular phantom colon, including fidicuary markers and haustral folds. The elimination of specular reflection was observed in the animal model, as ElasticFusion was able to track and recolor the surfels that changed color intensities in the video sequence. Quantitative validation is still work-in-progress, as the accuracy of the alignment of the 3D reconstruction needs to be further compared with other dense SLAM systems using our datasets and models.

Unlike other SLAM systems used in endoscopy, our framework uses a deep learning approach to estimate depth from a monocular video sequence, a Direct SLAM approach that is more robust to specular reflection and the sparsity of distinguishing features in endoscopic images, and requires no hardware modification. We also present an initial baseline for creating dense reconstructions in the GI tract, with our code and data planning to be made available.

The implications of reconstructing dense colon maps are numerous. Procedure statistics such as Cecal Intubation Rate (CIR), Adenoma Detection Rate (ADR), and withdrawal time are measured as key performance indicators [5], but the mosaicing approach outlined in this paper may provide much more relevant information about examination quality. For example, a withdrawal time (the time spent retracting the scope from the proximal colon to search for polyps) >6 minutes is often used as a key indicator of procedure quality [15], but what may actually be more informative is a report of areas canvased by the endoscopist, perhaps in units of fractional area captured. Experts in gastroenterology believe that in order for colonoscopy to be effective, approximately 90-95\% of the colon's surface should be inspected during the final withdraw phase of the procedure [16]. To the contrary, a review of procedures conducted by 65 endoscopist revealed that on average, only about 81\% of the mucosa is examined during screenings [17]. Suspected lesion management may also benefit from an aggregated reconstruction, allowing for easy relocalization of tissue sites requiring follow-up examination in future procedures. Current practice calls for the tattooing of the tissue site [18], but patient's run the risk of an inflammatory reaction or scarring of the colon tissue [5], making later resection more challenging.

Future work should focus on continuing to improve upon the accuracy of the reconstruction, with additional work on the depth estimation component of the input frames. The network presented in this work was trained to estimate a pseudo-depth with no absolute units, which leads inconsistencies in estimated geometric pose. A similar network may be trained with cinematic renderings and ground truth depth from CT in order to have a more accurate depth. Additionally, further research may focus on computing missed areas in the colon, such as those behind the semilunar folds of the large intestine, in order to bring these to the attention of the examining endoscopist. Missed regions may be determined by framing the cylindrical colon as a closed form model, with non-closed areas solved for to determine unobserved areas.

\section{Conclusion}

In summary, monocular images paired with depth estimation from a convolutional neural network show promise for improving the 3D reconstruction of endoscopic scenes. We present a depth estimation architecture that is robust to patient-specific colonic features by training with cinematically rendered colon images with domain randomization. These experiments demonstrate the capability to overcome the unique challenges of widefield endoscopic imaging, producing dense surface models of colonic features.

\section{Acknowledgements}
This work was supported in part with funding from the NIH NIBIB Trailblazer Award (R21 EB024700) and a sponsored research agreement with Olympus Medical.